# Measurement of the Boltzmann constant by the Doppler broadening technique at a $3.8 \times 10^{-5}$ accuracy level

K. Djerroud, C. Lemarchand, A. Gauguet, C. Daussy, S. Briaudeau♣, B. Darquié, O. Lopez, A. Amy-Klein, C. Chardonnet and Ch. J. Bordé

*Laboratoire de Physique des Lasers, UMR 7538 CNRS, Université Paris 13, 99 av. J.-B. Clément, 93430 Villetaneuse, France*

♣ *Institut national de Métrologie LNE-INM – CNAM – 62 rue du Landy, 93210 La Plaine Saint-Denis, France*

## Abstract

In this paper, we describe an experiment performed at the Laboratoire de Physique des Lasers and dedicated to an optical measurement of the Boltzmann constant $k_B$. With the proposed innovative technique, determining $k_B$ comes down to an ordinary frequency measurement. The method consists in measuring as accurately as possible the Doppler absorption profile of a rovibrational line of ammonia in thermal equilibrium. This profile is related to the Maxwell-Boltzmann molecular velocity distribution along the laser beam. A fit of the absorption line shape leads to a determination of the Doppler width proportional to $\sqrt{k_B T}$ and thus to a determination of the Boltzmann constant. The laser source is an ultra-stable $CO_2$ laser with a wavelength $\lambda \approx 10\,\mu m$. The absorption cell is placed in a thermostat keeping the temperature at 273.15 K within 1.4 mK. We were able to measure $k_B$ with a relative uncertainty as small as $3.8 \times 10^{-5}$, which represents an improvement of an order of magnitude for an integration time comparable to our previous measurement published in 2007 [1].

## Résumé

Dans cet article, nous présentons l'expérience développée au Laboratoire de Physique des Lasers pour la mesure optique de la constante de Boltzmann $k_B$. Cette nouvelle approche ramène la détermination de $k_B$ à une mesure de fréquence. L'expérience consiste à mesurer le plus exactement possible le profil d'absorption Doppler d'une raie de vibration-rotation de l'ammoniac à l'équilibre thermodynamique. Ce profil reflète la distribution de Maxwell-Boltzmann des vitesses moléculaires le long du faisceau laser. Une analyse de la forme de la raie d'absorption conduit à une détermination de l'élargissement Doppler, proportionnel à $\sqrt{k_B T}$, et donc à une mesure de la constante de Boltzmann. La mesure spectroscopique est réalisée à l'aide d'un laser à $CO_2$ ultra-stable de longueur d'onde $\lambda \approx 10\,\mu m$. La cellule d'absorption est placée dans un thermostat qui permet de contrôler la température autour de 273,15 K avec une incertitude de 1,4 mK. Ces mesures nous ont conduit récemment à une détermination de $k_B$ avec une incertitude relative de $3,8 \times 10^{-5}$. Cela représente, pour des temps de mesure comparables, un gain d'un ordre de grandeur par rapport à notre précédente mesure publiée en 2007 [1].

## Introduction

The Committee on Data for Science and Technology (CODATA) value for the Boltzmann constant $k_B$ essentially relies on a single experiment by Moldover *et al.* performed before 1988 and based on acoustic gas thermometry [2, 3]. The current relative uncertainty on $k_B$ is $1.7 \times 10^{-6}$ [4]. Since 1988, several projects have been developed to provide other measurements of this constant. Besides some projects following Moldover's approach [5-7], an alternative approach based on the virial expansion of the Clausius-Mossotti equation and measurement of the permittivity of helium is very promising [8]. This method has reached an uncertainty below 10 ppm and is taken into account by the CODATA. The renewed interest in the Boltzmann constant is related to the possible redefinition of the International System of Units (SI) expected to happen at one of the next meeting of the Conférence Générale des Poids et Mesures (CGPM) [9-16]. A new definition of the kelvin would fix the value of the Boltzmann constant to its value determined by CODATA.

In 2000, Ch. J. Bordé proposed a new approach for measuring the Boltzmann constant based on laser spectroscopy [17, 18]. With this method determining the Boltzmann constant comes down to a frequency measurement which is the physical quantity that can be measured with the highest accuracy. The principle is to record the Doppler profile of a well-isolated absorption line of an atomic or molecular gas in thermal equilibrium in a cell. This profile reflects the Maxwell-Boltzmann distribution of the velocity distribution along the laser beam axis. A complete analysis of the line shape which can take into account pressure broadening, optical depth, hyperfine structure, Lamb-Dicke-Mossbauer (LDM) narrowing, saturation of the molecular transition, *etc.* leads to a determination of the Doppler broadening and thus to $k_B$. In the Doppler limit, that is when all the above effects are negligible, the e-fold half-width of the Doppler profile, $\Delta\nu_D$, is given by: $\frac{\Delta\nu_D}{\nu_0} = \sqrt{\frac{2 k_B T}{mc^2}}$, where $\nu_0$ is the frequency of the molecular line, $c$ is the velocity of light, $T$ is the temperature of the gas and $m$ is the molecular mass. This mass is given by the ratio $M/N_A$ where $M$ is the molar mass and $N_A$ the Avogadro constant. Uncertainty on $k_B$ is then deduced from that on $T$, $m$ and $\Delta\nu_D$. The $m$ relative uncertainty is currently limited by that on the Avogadro constant to $2.9 \times 10^{-7}$ [4] (resulting from the silicon sphere experiment). Since atom interferometry yields a direct determination of the quantity $h/mc^2$ at a $10^{-8}$ level through a frequency measurement [19, 20], $m$ can also be determined knowing the Planck constant $h$. Since $h$ is measured by the watt balance experiment with an uncertainty of $5 \times 10^{-8}$ [4], this alternative approach leads to a better determination of $m$. One should then note that rewriting the previous formula as: $\frac{\Delta\nu_D}{\nu_0} = \sqrt{\frac{2(k_B/h)T}{(m/h)c^2}}$, brings to the fore the ratio $k_B/h$ whose uncertainty is limited by that on $T$, $mc^2/h$ and $\Delta\nu_D$. In the context of a redefinition of the SI, it becomes clear that fixing the value of $k_B/h$ would directly connect the temperature unit to the frequency unit.

We have recently developed a first experiment to demonstrate the potentiality of this new approach [1, 21]. The probed line is the $\nu_2$ saQ(6,3) rovibrational line of the ammonia molecule $^{14}NH_3$ at the frequency $\nu$ = 28 953 694 MHz. The choice of such a molecule is governed by two main reasons: a strong absorption band in the 8-12 μm spectral region of the ultra-stable spectrometer that we have developed for several years and a well-isolated Doppler line to avoid any overlap with neighbouring lines. Very encouraging preliminary results have led to a value of the Boltzmann constant with an uncertainty of 190 ppm [1]. Following these firsts results, at least three other groups have started to develop similar experiments on $CO_2$, $H_2O$ and acetylene [22-24]. In this paper, we present recent developments of our experimental set-up. A new approach to the data processing using a Voigt line shape will be described and our last measurement of $k_B$ will be detailed.

## I. Absorption line shape

The measurement principle is based on the linear absorption of a laser beam through a molecular vapor in a cell in thermodynamical equilibrium. An accurate determination of the Doppler line width requires a very accurate description of the line shape. We consider the case of an optically thick medium in which the absorption is given by the Beer-Lambert law $\exp(-\alpha L)$, where $L$ is the length of the absorption cell and $\alpha$ is the absorption coefficient. The absorption coefficient $\alpha$ can be described by a Voigt profile which is the convolution of a Gaussian related to the inhomogeneous Doppler broadening and a Lorentzian whose half-width, $\gamma_{hom}$, is the sum of all the homogeneous broadening contributions. Since the natural width is negligible for the rovibrational levels, this homogeneous width is dominated by the molecular collisions and, thus, is proportional to the pressure. In linear absorption spectroscopy and for an isotropic distribution of molecular velocities it can be shown that all transit effects are already included in the inhomogeneous Doppler broadening [25]. At high pressures, the LDM effect which results in a reduction of the Doppler width with pressure must be taken into account [26-29]. Finally, in our experiment two other effects that tend to broaden the line have to be considered: the unresolved hyperfine structure of the molecular transition and the modulations applied to the laser for detection purposes. These two effects come in to play only at second order and in any case can be calculated. Moreover, when molecules are not in a weak laser field regime, an additional saturation broadening may occur and should be avoided.

## II.  Earlier results and data analysis

In this section, we present the spectrometer, the data processing method and enumerate the main limitations of our earlier measurements.

### A.  *Experimental set-up*

The high accuracy laser spectrometer developed for decades in our group is used here in a specific optical set-up dedicated to the measurement of the Doppler broadened absorption spectrum of $NH_3$ around 29 THz (Figure 1). This measurement requires a fine control of: (i) the laser frequency which is tuned over a large frequency range to record the linear absoption spectrum; (ii) the laser intensity entering the absorption cell; (iii) the temperature of the gas which has to be measured during the experiment.

The spectrometer is based on a $CO_2$ laser and operates in the 8-12 μm range. For this experiment, the main issues for the $CO_2$ laser are its frequency stability, frequency tunability and intensity stability. The laser frequency stabilization scheme is described in reference [30]: a sideband generated with a tunable electro-optic modulator (EOM) is stabilized on an $OsO_4$ saturated absorption line detected on the transmission of a 1.6-m long Fabry-Perot cavity. The laser spectral width measured by the beat note between two independent lasers is smaller than 10 Hz and shows an Allan deviation of 0.1 Hz ($3.10^{-15}$) for a 100 s integration time.

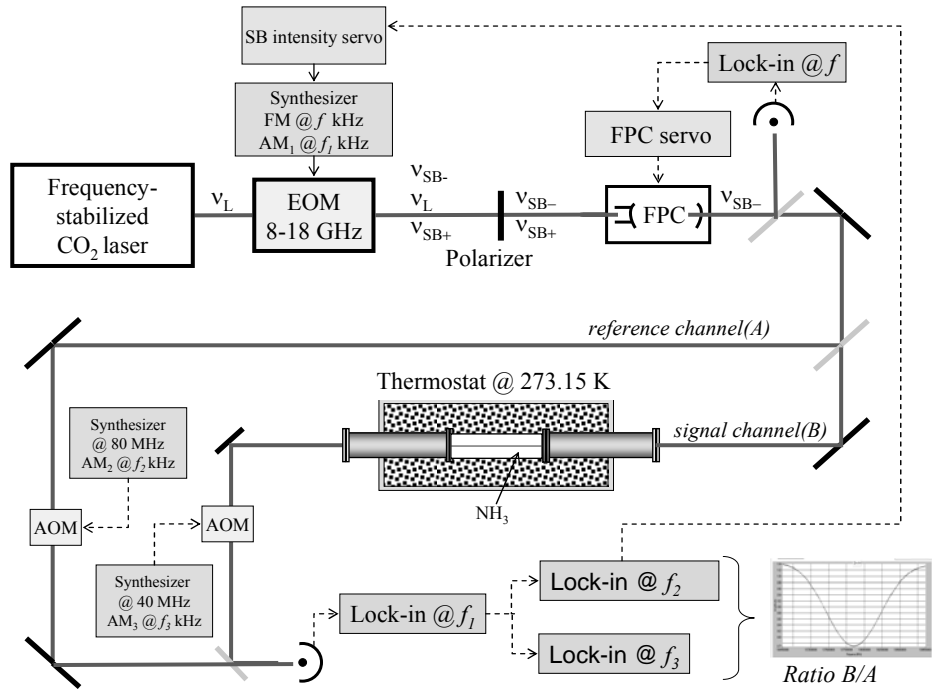

**Figure 1 : First generation experimental set-up (AM: amplitude modulation, FM: frequency modulation, EOM: electro-optic modulator, AOM: acousto-optic modulator, FPC: Fabry Perot cavity, SB: side-band, Lock-in: lock-in amplifier)**

Since its tunability is limited to 100 MHz, the CO$_2$ laser source is coupled to a second EOM which generates two sidebands of respective frequencies $\nu_{SB+} = \nu_L + \nu_{EOM}$ and $\nu_{SB-} = \nu_L - \nu_{EOM}$ on both sides of the fixed laser frequency $\nu_L$. The frequency $\nu_{EOM}$ is tunable from 8 to 18 GHz which is enough to scan and record the full Doppler profile. The intensity ratio between these two sidebands and the laser carrier is about 10$^{-4}$. After the EOM, a polarizer attenuates the carrier by a factor 200. A Fabry-Perot cavity (FPC) with a 1 GHz free spectral range and a finesse of 150 is then used to drastically filter out the residual carrier and the unwanted sideband $\nu_{SB+}$. The sidebands are frequency modulated via the EOM at a frequency $f$. Part of the transmitted signal from the FPC is detected at $f$ and used to lock its resonance frequency at the frequency $\nu_L - \nu_{EOM}$. In order to keep the laser intensity constant at the entrance of the cell during the whole experiment, the transmitted beam is then split in two parts with a 50/50 beamsplitter: one part feeds a 37-cm long ammonia absorption cell for spectroscopy (probe beam *B*) while the other is used as a reference beam (reference beam *A*). A detailed description of the detection scheme and intensity stabilization is given in reference [31]. The signal from channel *A* is permanently compared and locked to a very stable voltage reference. Thus, the signal *B* gives exactly the absorption signal of the molecular gas recorded with a constant incident laser power. In order to reject the laser carrier residual, the sideband is amplitude-modulated at a frequency $f_1$ via the EOM and the signal is detected at $f_1$. Reference and probe beams are detected on a single HgCdTe photodetector and amplitude-modulated at two different frequencies $f_2$ and $f_3$ by two acousto-optic modulators (AOM). The sideband is tuned close to resonance with the desired molecular transition and scanned to record the full Doppler profile by detecting the light transmitted through the cell. The gas pressure range in the cell is 1 – 8 Pa. The absorption profile, whose width is about 100 MHz, is recorded over 250 MHz by steps of 500 kHz with a 20 ms time constant. The time required to record a single spectrum is about 110 s, limited by the time delay for the FPC stabilisation.

The absorption cell is located inside a large thermostat filled with an ice-water mixture, which sets its temperature at 273.15 K. The thermostat is a large stainless steel box $(1 \times 0.5 \times 0.5 \, m^3)$ thermally isolated by a 10-cm thick insulating wall. The absorption cell placed at the center of the thermostat is a stainless steel vacuum chamber closed at each end with anti-reflective coated ZnSe windows. From these windows, pumped buffer pipes extending out of the thermostat walls are closed on the other side with room temperature ZnSe windows. The cell temperature and gradients are measured with several temperature probes, 100 Ω platinum resistors calibrated at the triple-point of water (TPW). The temperature control of the cell is better than 20 mK (70 ppm in relative value), only limited by the uncertainty of the temperature measurement.

## *B.   Data processing and first results*

In this section we describe the successive steps of the data analysis. Even if some theoretical line shape models are taking into account the LDM effect, such profiles could not be directly used to fit experimental spectra. Both the too low signal-to-noise ratio and the too high number of model parameters prevented a usual fitting algorithm from converging. Thus, numerical fits were performed with simpler profiles with a reduced number of independant parameters. As detailed further, although we first fitted the data with a simple exponential of a Gaussian, we then developed a more refined procedure which allowed us to use a Voigt profile. This method will be applied in the last section of this paper to our recent data. The LDM effect in our pressure conditions should then appear in the residuals.

In a first analysis, since the signal-to-noise ratio was not high enough to directly fit the data with the exponential of a Voigt profile, we tried to work as close as possible to the Doppler limit and fitted the experimental spectra with the exponential of a Gaussian. The fits were giving the two key parameters: amplitude and width of the Gaussian. As the pressure decreased towards zero, the line width was expected to converge to the Doppler width. Plotting the Gaussian width as a function of pressure, we thus estimated the Doppler width as the y-intercept of a linear fit of this data. The pressure scale was given by the amplitude of absorption determined from the fit. A set of 2000 spectra, with pressures ranging from 1 to 8 Pa, recorded over more than one month was used to mesure the Doppler e-fold half-width. The linear regression led to the following determination of the Doppler width:

$$\Delta\nu_D = 49.8831\,(47) \text{ MHz} \quad (9.5 \times 10^{-5})$$

which corresponds to the following measurement of the Boltzmann constant by laser spectroscopy:

$$k_B = 1.380\,65\,(26) \times 10^{-23} \text{ J K}^{-1}$$

with a statistical uncertainty of 190 ppm [1]. This uncertainty was mainly limited by the signal-to-noise ratio of 170 for a time constant of 20 ms. The laser frequency instability had negligible impact on the uncertainty of $k_B$ (less than $10^{-7}$). This value is in good agreement with the CODATA value.

A detailed analysis of parasitic effects which can be responsible for a systematic error in the Doppler width determination showed that under our experimental conditions, the main systematic error was due to the oversimplified line shape used to fit the spectra. As the pressure decreased towards zero, we verified that the experimental line shape converged towards the exponential of a Gaussian profile, by checking that the residuals became negligible. However as the pressure increased, this simplified profile became less and less suitable to reproduce spectra and larger residuals were observed (Figure 2).

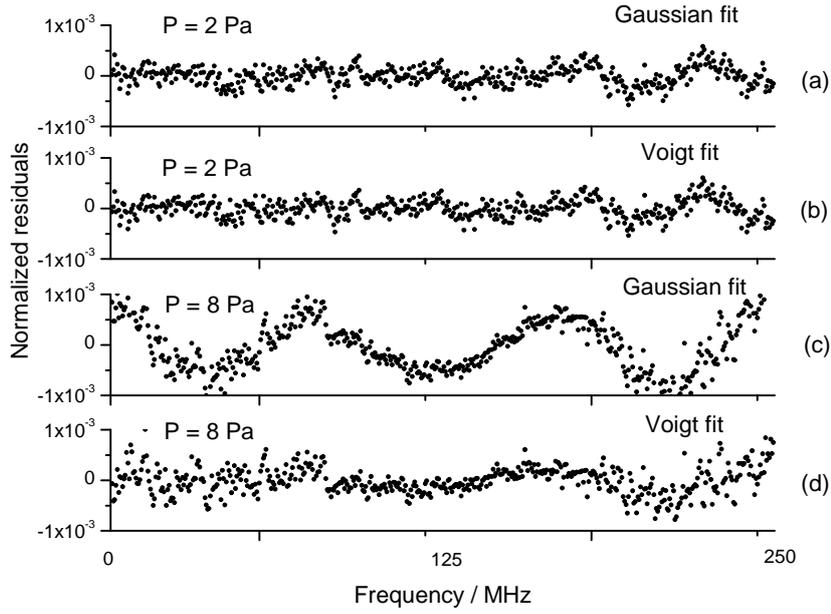

**Figure 2 : Normalized residuals (averaged over 30 spectra) for a non-linear least-squares fit at 2 and 8 Pa, with either the exponential of a Gaussian ((a) and (c)), or the exponential of a Voigt profile ((b) and (d)). The frequency scale is offset by 28 953 694 MHz**

In order to estimate precisely the impact of the simplifed Gaussian profile used to fit the data, we generated a set of exponential of a Voigt profile, for gas pressures ranging from 1 to 8 Pa, corresponding to our experimental conditions. These simulated profiles were then fitted with the exponential of a Gaussian and exactly the same procedure was implemented as described above for the experimental data. This enabled us to estimate that the Doppler width was biased by a systematic error of **-180 ppm**.

As a consequence, we decided to develop an improved procedure based on a Voigt profile which made it possible to determine independently both the homogeneous and the Doppler width. As mentioned in section I, a Voigt profile is the convolution of a Gaussian (e-fold half-width $\Delta \nu_D$) and a Lorentzian whose half-width, $\gamma_{hom} = \gamma P$ is, in our conditions, the pressure broadening (*P* being the pressure). Thus the collisional broadening $\gamma$ is a parameter shared by all the spectra, whatever the pressure *P* is, while the latter is proportional to the amplitude of absorption spectrum as mentioned earlier. We know that the signal-to-noise ratio of individual spectra is not good enough to disentangle properly the homogeneous and the Doppler width from the Voigt profile. Thus, to make the fitting algorithm converge, we decided to first guess an initial realistic value for $\gamma$. The improved fitting procedure is thus as follows. We fit all the experimental spectra with a Voigt profile, constraining $\gamma$ to its guessed value, leaving only $\Delta \nu_D$, *P* (both in the amplitude and $\gamma_{hom}$), the resonance frequency of the absorption line, and the baseline as adjustable parameters. We expect $\Delta \nu_D$ to remain constant when the pressure varies, if $\gamma$ is chosen equal to the correct value, but this is unlikely to happen at the first iteration. As in the previous procedure, we then plot $\Delta \nu_D$ as a function of *P* and record the slope *s* given by a linear regression of this data. We repeat this procedure for different values of $\gamma$ leading both to negative and positive slopes. A plot of *s* versus $\gamma$ shows a linear behaviour and a linear regression enables us to find the correct value of $\gamma$ for which *s* is zero. The experimental data are finally fitted again constraining $\gamma$ to its estimated final value for which $\Delta\nu_D$ remains constant (within the noise) whereas the pressure varies. A weighted average of all the $\Delta\nu_D$ gives the best estimate of the Doppler width from which we deduce the Boltzmann constant. One should note that with this technique $\gamma$ is adjusted in such a way that it is constrained to a constant value for all the measured spectra. Following this new fitting procedure for the same set of 2000 spectra previously processed with the first crude method, we obtained the following value for the Boltzmann constant: $1.380\,880(99) \times 10^{-23}\,\text{JK}^{-1}$ with a noticeably reduced statistical uncertainty of **72 ppm**, limited by the signal-to-noise ratio. At the same time, it is obvious from Figure 2 that fitting with the exponential of a Voigt profile gives better residuals compared to a Gaussian profile.

Nevertheless, as displayed in Figure 2, at the higher pressures, even with a Voigt profile, residuals seem to show some structures. The profile used for the adjustement can be further refined by taking into account the narrowing of the line shape at higher pressure by the LDM effect [26]. Various theoretical models are available in the literature, depending on the assumption made for the type of collisions between molecules [32]. Such new profiles need at least one additional independent parameter to describe the line shape. Among them the Galatry profile [27] makes the assumption of so-called "soft" collisions between molecules with the introduction of the diffusion coefficient as a new parameter. In order to estimate the systematic error made on our above estimate of $k_B$ due to the LDM effect (not included in the Voigt profile used up to this point), we generated numerically a set of Galatry profiles, for gas pressures ranging from 1 to 8 Pa, corresponding to our experimental conditions. The self-diffusion coefficient was chosen to be $D_{NH_3} = 0.16\ cm^2 s^{-1}$ according to measurements and calculations found in the literature [33, 34]. We then fitted these simulated profiles with a Voigt profile, using the exact same procedure as described earlier for the experimental data. This enabled us to demonstrate that the systematic error on $k_B$ due to the LDM effect (within the "soft" collisions assumption) is of about -90 ppm for pressures ranging from 1 to 8 Pa and can be reduced below **-28 ppm** by recording spectra of ammonia at pressures lower than 1.3 Pa.

## III. Second generation experimental set-up

These first and preliminary results are very encouraging and have led to a new set-up with a new cell, which can be used in either single or multiple path configuration, a new temperature control and a new detection scheme that we describe in this section.

After a thorough analysis of noise sources in our experiment, the detection scheme has been modified: two photodetectors are now used, one for the signal channel and one for the reference channel and the modulation scheme, greatly simplified, avoids frequency modulation which was responsible for extra-broadening and excess noise. The new set-up is shown in Figure 3. The $CO_2$ laser is stabilized as described earlier for the first series of experiments. The reference beam (*A*) and the probe beam (*B*) which crosses the absorption cell are amplitude-modulated at *f* = 44 kHz via the 8-18 GHz EOM and signals are recovered with a demodulation at *f*. The reference signal *A* which gives the intensity of the sideband $\nu_L - \nu_{EOM}$ is compared and locked to a very stable voltage reference by acting on the length of the FPC. The probe beam (*B*) signal then gives exactly the absorption signal of the molecular gas recorded with a constant incident laser power governed by the stabilization of signal *A*. The frequency servo of the filtering FPC as well as the intensity servo of the $CO_2$ laser have been reconsidered and simplified. The number of feedback loops has been reduced from two to one. In this set-up, the number of modulations applied to the laser beam was reduced from 3 amplitude modulations and 1 frequency modulation to only 1 amplitude modulation. All these modifications led to a reduction of detection noise by a factor 3 and a reduction of the single spectrum acquisition duration by a factor 3. Analysis of the former 2000 spectra revealed a very tiny systematic effect due to stray light reaching the detector. Optimization of the optical alignment to reach better optical isolation, as well as laser beam spatial filtering with a pinhole led to a large reduction, by a factor 30, of the residual fluctuations observed on the profile base line.

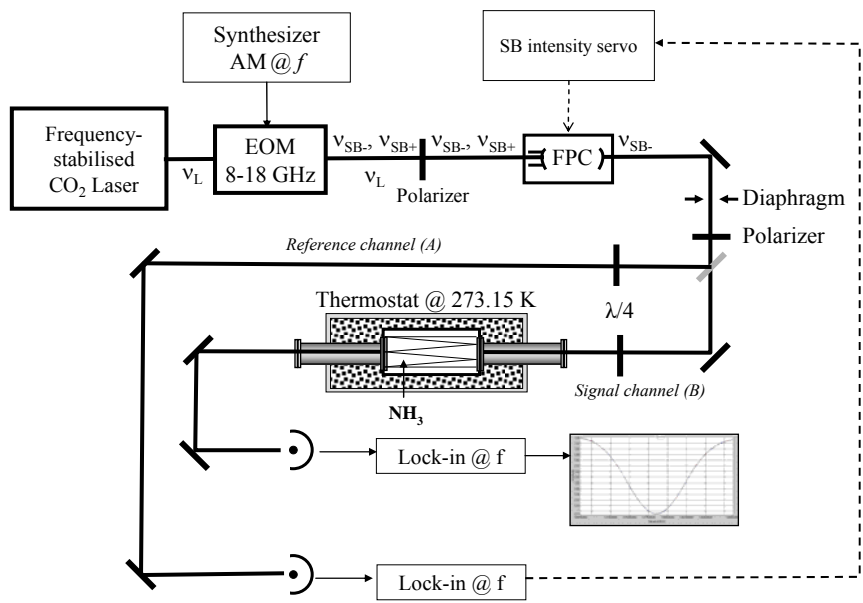

**Figure 3: Second generation experimental set-up (AM: amplitude modulation, EOM: electro-optic modulator, FPC: Fabry Perot cavity, SB: side-band, Lock-in: lock-in amplifier)**

In addition, a new cell of adjustable length was built and can either be used in a 37-cm long single path or a 3.5-m long multiple path configuration (Figure 4). Such a path length of 3.5 m was chosen in order to record spectra at lower pressures and thus reduce the LDM effect associated systematic error below 28 ppm, as explained in the previous section.

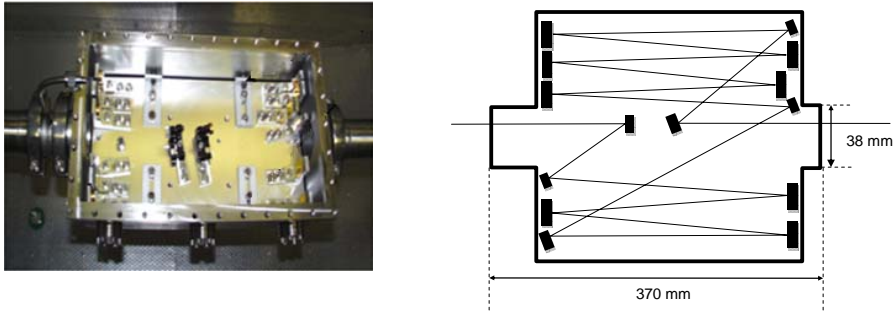

**Figure 4 : Multiple path cell : optical sketch (right) and picture (left)**

Since the absorption signal amplitude is proportional to the absorption length, the multiple path configuration leads to an improvement of the signal-to-noise ratio by a factor 10 at low pressure. Examples of spectra obtained for pressures ranging from 0.4 Pa to 5 Pa are displayed in Figure 5.

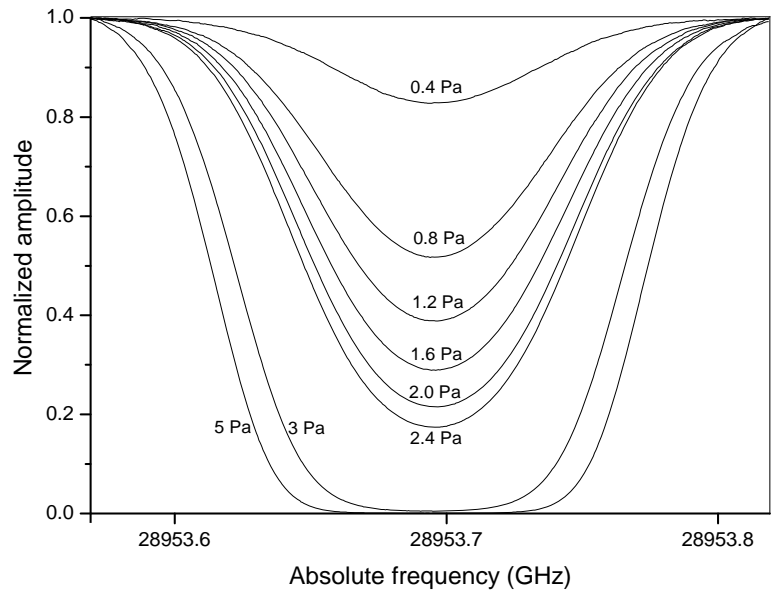

**Figure 5: Spectra recorded in the 3.5-m long multiple path configuration and for pressures ranging from 0.4 to 5 Pa**

The absorption cell is immersed in an ice-water mixture (Figure 6). Its temperature is measured with 3 precision 25 Ω Standard Platinum Resistance Thermometers (SPRTs) calibrated at the TPW and at the gallium melting point. The SPRTs are coupled to a new resistance measuring bridge (Guildline Instruments Limited, ref 6675/A) calibrated versus a resistance standard with a very low temperature dependence.

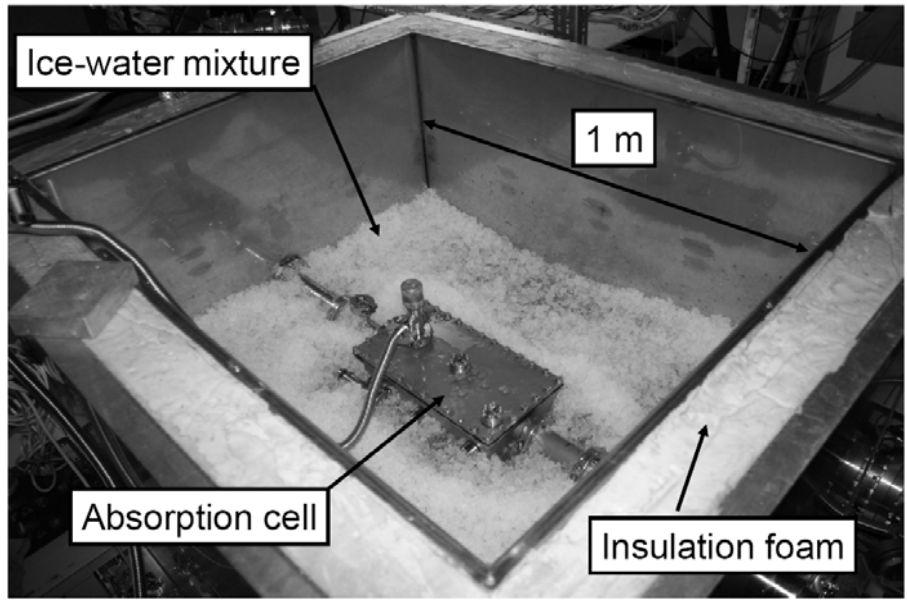

**Figure 6 : Multiple path absorption cell inside the ice-water thermostat**

As a consequence, the temperature accuracy is 1 ppm with a noise of 0.2 ppm for 40 s integration time. For longer integration times, drifts of the temperature remain below 1 ppm/day. A reproducible residual gradient is measured along the absorption cell. The vertical (resp. horizontal) gradient is equal to 0.05 mK/cm (resp. 0.03 mK/cm) leading to a global inhomogeneity of temperature along the cell below **5 ppm**.

## IV. Recent results

This new experimental set-up was used to record a series of spectra for pressures ranging from 0.1 to 1.3 Pa in the multiple path configuration. Absorption profiles were recorded over 250 MHz by steps of 500 kHz with a 30 ms time constant per point. For a 100% absorption, the signal-to-noise ratio was typically $10^3$. The time needed for recording one spectrum was reduced to 42 s. The Doppler widths obtained after processing 1420 spectra recorded at various pressures are presented in Figure 7. The improved fitting procedure based on a Voigt profile and detailed in section II.B was used here.

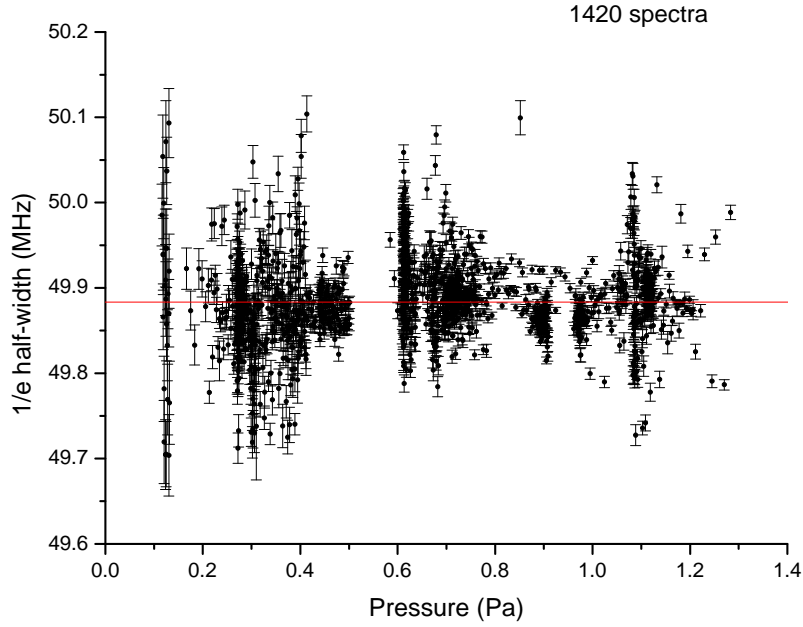

**Figure 7: e-fold half-width of the Gaussian contribution to the sa Q(6,3) NH$_3$ absorption line versus pressure for 1420 measurements, after fitting spectra with a Voigt profile (see section II.B)**

After 16 hours of accumulation, 1420 spectra yielded a statistical uncertainty on $k_B$ reduced to 38 ppm, limited by noise detection. When compared to the result published in 2007 [1], this represents an order of magnitude improvement for comparable measurement times.

A weighted average of all the data points displayed in Figure 7 led to a Doppler width of:
$$\Delta\nu_D = 49.88339 \ (93) \ \text{MHz} \ (1.9 \times 10^{-5})$$
The corresponding value for the Boltzmann constant is:
$$k_B = 1.380\ 669\ (52) \times 10^{-23} \text{JK}^{-1} \ (3.8 \times 10^{-5})$$

This value was not corrected for any systematic shift and the given uncertainty of 38 ppm is only of statistical origin. Attempts to observe systematic effects due to the modulation index, to the size or the shape of the laser beam, and to the laser power, including a non-linearity of the photodetector were unsuccessful at the 10 ppm level. As explained in detail in section II.B, we demonstrated that the systematic shift due to the LDM effect (Galatry profile) is below 28 ppm for pressures lower than 1.3 Pa. We believe that if we had assumed another collisional model this result would have been comparable. We had a careful look at the residuals and checked that that they became negligible at low pressures which indicates that the experimental line shape converged well towards the exponential of a Voigt profile. Finally no systematic effects due to the temperature control of the cell have been observed.

# V. Conclusion

After a very promising first demonstration of an optical measurement of the Boltzmann constant which reached an uncertainty of 190 ppm, a second generation experimental set-up was designed and implemented. **The value of the Boltzmann constant deduced from a second series of experiments performed on the new set-up is equal to $1.380\,669(52) \times 10^{-23}$ JK$^{-1}$. The statistical uncertainty is 38 ppm. This measurement is in agreement with the value recommended by CODATA,** $1.380\,6504(24) \times 10^{-23}$ JK$^{-1}$ [4], **within 13 ppm.** Improvement of the statistical uncertainty of the Boltzmann constant measurement over the years at the Laboratoire de Physique des Lasers is illustrated in Figure 8.

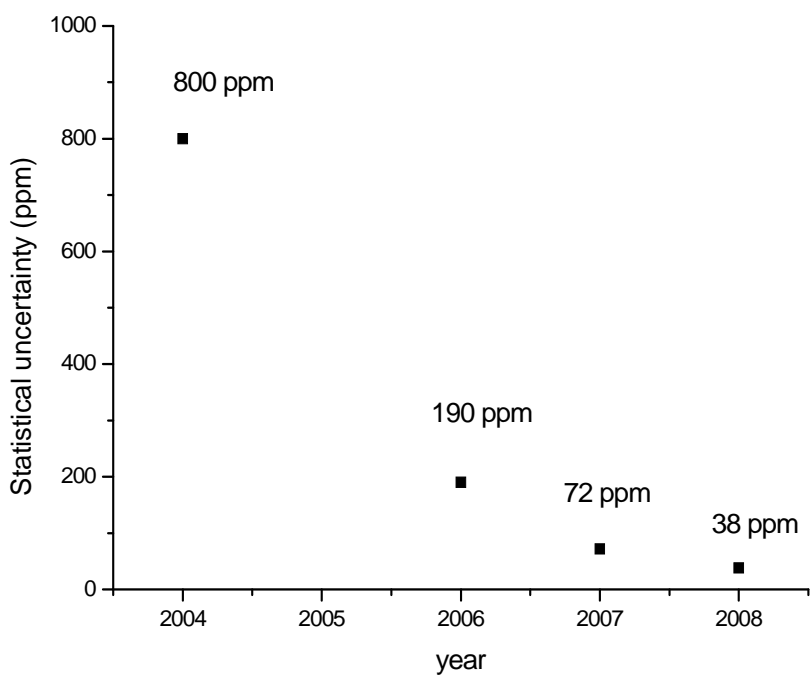

**Figure 8: Statistical uncertainty of the Boltzmann constant measurements at the Laboratoire de Physique des Lasers from 2004 until 2008 [1, 21, 35]**

To reduce this uncertainty, the thermostat needs to be strongly improved by using several insulator/conductor layers, which should guarantee a stability and an homogeneity of the temperature better than 1 ppm. A new multiple path Herriott cell (37 m absorption length) will be installed on the spectrometer to record spectra at even lower pressures for which we expect that systematic shifts due to velocity-changing collisions can be neglected at the 1 ppm level. These straightforward directions of progress let us expect a gain of 1 order of magnitude in the measurement uncertainty. The high sensitivy of our spectrometer will provide the conditions needed to study quantitatively all systematic effects at a few ppm level.

A study of line shape profiles at higher pressures is also under progress. A new 3 cm-long absorption cell has been designed to record spectra at pressures up to 200 Pa. As a consequence, this experiment will provide very accurate absorption profiles on a large pressure range (more than 4 orders of magnitude) readily relevant to molecular collisions and atmospheric physics studies. Such data will be used to analyse line shape profiles in the LDM regime. A detailed study of molecular diffusion effects and collisional relaxation rates will be performed. We will be able to test, in our experimental conditions, the relevance of various theoretical models that include the velocity dependence of relaxation rates, such as the Galatry profile ("soft" collisions assumption [24]) or the Rautian profile ("hard" collisions assumption [29], actually introduced for the first time in spectroscopy by M. Nelkin and A. Ghatak in 1964 [28]).

Finally this project could lead to the development of an absolute thermometer operating at high temperature up to several hundreds of degrees Celsius. In this range the accuracy of our experiment should exceed that of the current techniques. This is very appealing as an approach for improving the temperature scale.

**Acknowledgements:**

This work is funded by the Laboratoire National de Métrologie et d'Essais and by European Community (EraNet/IMERA). Authors would like to thank Y. Hermier, F. Sparasci and L. Pitre from LNE-INM/CNAM for SPRTs calibrations, discussions and advice for the thermometry part of this project.